\documentclass[aps,reprint,twocolumn,floatfix,showpacs,superscriptaddress,pra]{revtex4-1}
\usepackage[colorlinks,linkcolor=blue,anchorcolor=blue,citecolor=blue,filecolor=blue,menucolor=blue,runcolor=blue,urlcolor=blue,frenchlinks=blue]{hyperref}
\usepackage{graphicx}
\usepackage{subfigure}
\usepackage{amsmath}
\usepackage{amssymb}
\usepackage{dcolumn}
\usepackage{mathrsfs}
\usepackage{color,amsfonts,amsthm,mathtools}
\usepackage{subfigure}
\usepackage{setspace}
\usepackage{bm}

\makeatletter

\newcommand{\Rmnum}[1]{\expandafter\@slowromancap\romannumeral #1@}
\makeatother

\begin{document}
\title{Non-Abelian quantum geometric tensor in degenerate topological semimetals}

\author{Hai-Tao Ding}
\affiliation{National Laboratory of Solid State Microstructures and School of Physics, Nanjing University, Nanjing 210093, China}
\affiliation{Collaborative Innovation Center of Advanced Microstructures, Nanjing 210093, China}
\affiliation{Department of Physics, National University of Singapore, Singapore 117551}

\author{Chang-Xiao Zhang}
\affiliation{National Laboratory of Solid State Microstructures and School of Physics, Nanjing University, Nanjing 210093, China}
\affiliation{Collaborative Innovation Center of Advanced Microstructures, Nanjing 210093, China}

\author{Jing-Xin Liu}
\affiliation{National Laboratory of Solid State Microstructures and School of Physics, Nanjing University, Nanjing 210093, China}
\affiliation{Collaborative Innovation Center of Advanced Microstructures, Nanjing 210093, China}

\author{Jian-Te Wang}
\affiliation{National Laboratory of Solid State Microstructures and School of Physics, Nanjing University, Nanjing 210093, China}
\affiliation{Collaborative Innovation Center of Advanced Microstructures, Nanjing 210093, China}

\author{Dan-Wei Zhang}
\email{danweizhang@m.scnu.edu.cn}
\affiliation{Key Laboratory of Atomic and Subatomic Structure and Quantum Control (Ministry of Education), Guangdong Basic Research Center of Excellence for Structure and Fundamental Interactions of Matter, South China Normal University, Guangzhou 510006, China}
\affiliation{Guangdong Provincial Key Laboratory of Quantum Engineering and Quantum Materials, School of Physics, South China Normal University, Guangzhou 510006, China}

\author{Shi-Liang Zhu}
\email{slzhu@scnu.edu.cn}
\affiliation{Key Laboratory of Atomic and Subatomic Structure and Quantum Control (Ministry of Education), Guangdong Basic Research Center of Excellence for Structure and Fundamental Interactions of Matter, South China Normal University, Guangzhou 510006, China}
\affiliation{Guangdong-Hong Kong Joint Laboratory of Quantum Matter, Frontier Research Institute for Physics, School of Physics, South China Normal University, Guangzhou 510006, China}

\date{\today}

\begin{abstract}
The quantum geometric tensor (QGT) characterizes the complete geometric properties of quantum states, with the symmetric part being the quantum metric, and the antisymmetric part being the Berry curvature. We propose a generic Hamiltonian with global degenerate ground states, and give a general relation between the corresponding non-Abelian quantum metric and unit Bloch vector. This enables us to construct the relation between the non-Abelian quantum metric and Berry or Euler curvature. To be concrete, we present and study two topological semimetal models with global degenerate bands under $CP$ and $C_2T$ symmetries, respectively. The topological invariants of these two degenerate topological semimetals are the Chern number and Euler class, respectively, which are calculated from the non-Abelian quantum metric with our constructed relations. Based on the adiabatic perturbation theory, we further obtain the relation between the non-Abelian quantum metric and the energy fluctuation. Such a non-adiabatic effect can be used to extract the non-Abelian quantum metric, which is numerically demonstrated for the two models of degenerate topological semimetals. Finally, we discuss the quantum simulation of the model Hamiltonians with cold atoms. 
\end{abstract}
\maketitle

\section{Introduction}
Geometry and topology play central roles in various fields of modern physics, and broaden our horizons about the classification of quantum phases. A representative example of this is the discovery of the integer quantum Hall effect in the 1980s \cite{Klitzing1980}, which goes beyond the Landau's theory of phase transitions and can be understood as a topological effect characterized by the Thouless-Kohmoto-Nightingale-den Nijs invariant (also called Chern number) \cite{Thouless1982,Simon1983,Berry1984}. Since then, topological insulators and semimetals have been explored in condensed-matter physics and engineered systems \cite{Hasan2010,XLQi2011,Armitage2018,Cooper2019,DWZhang2018,Goldman2016}. Classification of topological quantum states in terms of antiunitary symmetries (e.g., time reversal symmetry $T$) and unitary symmetries (e.g., reflection symmetry and $C_{n}$ rotation symmetry) has been widely studied. For instance, the classification of topological semimetals was established with a unified theory in Ref. \cite{YXZhao2016}, when the systems have $PT$ ($P$ is inversion symmetry) and $CP$ symmetry-protected nodal band structures. 

The complete geometric properties of these Bloch functions are fully encoded by the quantum geometric tensor (QGT) \cite{Provost1980,Michael2017,Nakahara2003}. Its imaginary part is the Berry curvature \cite{Berry1984,DWZhang2018}, which is closely related to many interesting effects~\cite{MBerry1989,DXiao2010,XShen2018,XShen2019,XDHu2021,ZXGuo2022,XShen2022,SLZhu2013,YPWu2022,DWZhang2016}. The integral of the Berry curvature over a closed two-dimensional (2D) manifold defines the first Chern number. The real part of QGT is the so-called quantum metric, characterizing the distance between two nearby quantum states for both degenerate (non-Abelian) and non-degenerate (Abelian) systems \cite{Michael2017}. The QGT is related to various physical observables \cite{Neupert2013,Kolodrubetz2013,Albert2016,Bleu2018,OzawaT2018,Albuquerque2020,Lapa2019,YGao2019,Legner2013, Klees2020,TOzawa2021,CDGrandi2013,PHe2021,Gersdorff2021,RQWang2019,RQWang2021a,RQWang2021b,BHet2023}, and has been experimentally measured in engineered quantum systems \cite{XSTan2018,JZLi2022,QXLv2023,MLysne2023,QLiao2021,WZheng2022,CRYI2023,MYu2022,TanXS2019,MYu2020,XTan2021,MChen2022,Gianfrate2020,LAsteria2019,MYu202256}. However, most of these studies are limited to the Abelian QGT in non-degenerate systems.

In recent years, the non-Abelian QGT in degenerate quantum systems has been investigated, and some detection schemes based on dynamical responses have been proposed \cite{HTDing2022,HWeisbrich2021,WZheng2022}. If a system's Hamiltonian changes very slowly, and the system is prepared in one of the eigenstates of the system's Hamiltonian, then at time $t$, it will remain in the instantaneous eigenstate. This is known as the adiabatic theorem \cite{Messiah1962}. However, for the case that the Hamiltonian is not slowly enough varied, researchers have presented the adiabatic perturbation theory, i.e., perturbation theory in the instantaneous basis \cite{GRigolin2008,GRigolin2014,GRigolin2010,GRigolin2012}, to describe the evolution of the quantum state. For degenerate systems, when slowly ramping the system's parameters, in some cases,  response coefficients of the time-dependent quantum states can be related to the non-Abelian QGT \cite{Abigail2018,Kolodrubetz2016,Michael2017,Kolodrubetz2013}. For example, the non-Abelian Berry curvature related to the so-called generalized force and the corresponding second Chern number have been measured \cite{Abigail2018,Kolodrubetz2016}.

The aim of present work is twofold. On one hand, we propose a general Dirac Hamiltonian with global degenerate bands and establish relations between non-Abelian quantum metric and Berry curvature wherein. On the other hand, we show that the non-Abelian quantum metric corresponds to the energy fluctuation, acting as a non-adiabatic effect and providing an alternative method to extract all the components of non-Abelian QGT. Concretely, we study two three-dimensional (3D) lattice Hamiltonians with $CP$ and $C_{2}T$ symmetries \cite{YXZhao2016,ABouhon2020}, respectively. Within certain parameter ranges, the topological semimetal phases are exhibited with the monopole charges characterized by Chern number and Euler class \cite{Unal2020,ABouhon2020,MEzawa20211,ATimmel2023,ABouhon202020,Jankowski2023}, respectively. We find that these two topological invariants can be calculated from non-Abelian quantum metric. When one of the momenta is fixed as constant, the 3D models are reduced to 2D models with topological insulator phases and symbolic gapless boundary modes, which are characterized by corresponding topological invariants and the Wilson loops. We reveal the intrinsic relations between non-Abelian quantum metric and Berry/Euler curvature. These relations provide alternative methods to calculate the Chern number and Euler class. We also numerically demonstrate the measurements of the non-Abelian QGT with the non-adiabatic effect in the two models of degenerate topological semimetals. Furthermore, we discuss the quantum simulation of the model Hamiltonians in controlled ultracold atomic platforms \cite{Price2015,Lohse2018,Boada2012,Celi2014,Guglielmon2018,YLChen2020,DucaL2016,Eckardt2017,Reitter2017,Aidelsburger2015,FMei2014,GLiu2010,FMei2012,DWZhang2020}.

This paper is organized as follows. In Sec. \ref{sec2}, we give a brief review of the QGT and propose a generic Dirac Hamiltonian with global degenerate bands and intrinsic relations between the real and imaginary parts of its non-Abelian QGT. In Sec. \ref{sec3} and Sec. \ref{sec4}, we apply the relations to the concrete $CP$ symmetric and $C_2T$ symmetric Hamiltonians, respectively, and study the exotic properties of the degenerate topological semimetals. In Sec. \ref{sec5}, we derive the relation between non-Abelian quantum metric and the energy fluctuation from adiabatic perturbation theory, and numerically demonstrate the dynamic scheme to extract the non-Abelian QGT. The experimental schemes to simulate the two model Hamiltonians with ultracold atoms are proposed in Sec. \ref{sec6}. Finally in Sec. \ref{sec7}, a short conclusion is given.

\section{Non-Abelian quantum geometric tensor}
\label{sec2}
We consider a generic Hamiltonian $H(\boldsymbol\lambda)$ parameterized by $\boldsymbol\lambda=(\lambda_1,\lambda_2,...)$ with $N$ degenerate ground states $\left|\psi_{j}(\boldsymbol\lambda)\right\rangle$ $(j=1,2,...,N)$. One of the ground states can be expanded as $\left|\Psi_{0}(\boldsymbol\lambda)\right\rangle=\sum_{j=1}^{N} C_{j}(\boldsymbol\lambda\left|\psi_{j}(\boldsymbol\lambda)\right\rangle$, where $\sum_{j=1}^{N}|C_j(\boldsymbol\lambda)|^2=1$. Then the distance between two nearby quantum states \cite{Mayuquan2010,Michael2017}  $\left|\Psi_{0}(\boldsymbol\lambda)\right\rangle$ and  $\left|\Psi_{0}(\boldsymbol\lambda+d\boldsymbol\lambda)\right\rangle$ is defined as
\begin{equation}
\label{S}
\begin{aligned}
d S^{2} &=1- \left|\left\langle \Psi_{0}(\boldsymbol\lambda) \mid \Psi_{0}(\boldsymbol\lambda+d\boldsymbol\lambda) \right\rangle\right|^{2} \\
&=\sum_{\mu \nu}\left[\left(C_{1}^{*} \quad \cdots \quad C_{N}^{*}\right) Q_{\mu \nu}\left(\begin{array}{c}
C_{1} \\
\vdots \\
C_{N}
\end{array}\right)\right] d \lambda_{\mu} d \lambda_{\nu}  \\
&=\sum_{\mu \nu}\left[\left(C_{1}^{*} \quad \cdots \quad C_{N}^{*}\right) g_{\mu \nu}\left(\begin{array}{c}
C_{1} \\
\vdots \\
C_{N}
\end{array}\right)\right] d \lambda_{\mu} d \lambda_{\nu},
\end{aligned}
\end{equation}
where $Q_{\mu \nu}$ is a $N \times N$ matrix, with the matrix element
\begin{equation}
Q_{\mu \nu}^{i j}:=\left\langle\partial_{\mu} \psi_{i}(\boldsymbol\lambda)|[1-P(\boldsymbol\lambda)]| \partial_{\nu} \psi_{j}(\boldsymbol\lambda)\right\rangle.
\end{equation}
Here all derivatives are taken with respect to the parameters, i.e., $\partial_{\mu}=\partial_{\lambda_{\mu}}$, and $P(\boldsymbol\lambda)$ is a projection operator of ground  states
\begin{equation}
P(\boldsymbol\lambda)=\sum_{j=1}^{N}\left|\psi_{j}(\boldsymbol\lambda)\right\rangle\left\langle\psi_{j}(\boldsymbol\lambda)\right|.
\end{equation}
The corresponding non-Abelian quantum metric $g_{\mu \nu}$ and Berry curvature $F_{\mu \nu}$ are
\begin{equation}
\label{g}
\begin{aligned}
g_{\mu \nu} &=(Q_{\mu \nu}+Q_{\mu \nu}^{\dagger})/2, \\
F_{\mu \nu} &=i(Q_{\mu \nu}-Q_{\mu \nu}^{\dagger}),
\end{aligned}
\end{equation}
respectively. These two quantities are also $N\times N$ matrices, with $g_{\mu \nu}=g_{\mu \nu}^{\dagger}=g_{\nu \mu}$ and $F_{\mu \nu}=F_{\mu \nu}^{\dagger}=-F_{\nu \mu}$. The components of $g_{\mu\nu}$ and $F_{\mu\nu}$ are written as
\begin{equation}
\label{gf}
\begin{aligned}
g_{\mu\nu}^{ij}&=(Q_{\mu\nu}^{ij}+Q_{\nu\mu}^{ij})/2,  \\
F_{\mu\nu}^{ij}&=i(Q_{\mu\nu}^{ij}-Q_{\nu\mu}^{ij}).
\end{aligned}
\end{equation}
For the case of $N=1$, there is no degeneracy for ground state, and the corresponding non-Abelian QGT is simplified to the Abelian QGT. In the rest of this paper, we focus on the case of $N=2$.

To meet the symmetry requirement as discussed later, we consider a Dirac Hamiltonian with the following form
\begin{equation}
\label{H_gen}
H(\boldsymbol\lambda)=d_1 \Gamma_1 + d_2 \Gamma_2 +d_3 \Gamma_3,
\end{equation}
where $d_{i}$ $(i=1,2,3)$ are the components of Bloch vector $\textbf{d}=(d_1,d_2,d_3)$ and real functions parameterized by $\boldsymbol\lambda=(\lambda_{\mu},\lambda_{\nu})$, $\Gamma_i$ $(i=1,2,3)$ are $4 \times 4$ Clifford matrices that satisfy the anti-commutation relations $\left\{\Gamma_i,\Gamma_j\right\}=2\delta_{ij}$. Under proper conditions as given in Appendix \ref{appa}, the two valence bands of the Hamiltonian in Eq.~(\ref{H_gen}) are global degenerate across all parameter space, with the energy dispersion given by $E_{ \pm}=\pm |\textbf{d}|=\pm \sqrt{d_1^2+d_2^2+d_3^2}$.
We consider the subspace spanned by two degenerate ground states $\left\{\left|\psi_{1}(\boldsymbol\lambda)\right\rangle, \left|\psi_{2}(\boldsymbol\lambda)\right\rangle \right\}$,
\begin{equation}
\text{Tr} (g_{\mu\nu})=g_{\mu\nu}^{11}+g_{\mu\nu}^{22}=\frac{1}{2}(\partial_{\mu}\hat{\mathbf{d}} \cdot \partial_{\nu}\hat{\mathbf{d}}) ,  \\
\end{equation}
where the unit Bolch vector $\hat{\mathbf{d}}\equiv \textbf{d}/|\textbf{d}|=(\hat{d}_1,\hat{d}_2,\hat{d}_3)$. Using the method outlined in Ref. \cite{AZhang2022} and in Appendix \ref{appc}, we derive the following general relation
\begin{equation}
\label{gfre}
\sqrt{\operatorname{det} G_{\mu\nu}}  =\left|\epsilon_{\alpha \beta \gamma} \hat{d}_\alpha \partial_{\mu} \hat{d}_\beta \partial_{\nu} \hat{d}_\gamma\right|.
\end{equation}
where $\epsilon_{\alpha\beta\gamma}$ is the Levi-Civita symbol with $\left\{\alpha,\beta,\gamma\right\}=\left\{1,2,3\right\}$, and the matrix $G_{\mu\nu}$ has the form
\begin{equation}
 \frac{G_{\mu\nu}}{2}=\left(\begin{array}{ccc}
\text{Tr}(g_{\mu\mu}) & \text{Tr}(g_{\mu\nu})
\vspace{1ex}\\
\text{Tr}(g_{\nu\mu}) & \text{Tr}(g_{\nu\nu})  \\
\end{array}\right).                                   \\
\end{equation}
The relation in Eq.~(\ref{gfre}) can be easily extended to Hamiltonians composed of $2^{N}\times 2^{N}$ Clifford matrices. In a two-band model, the Abelian Berry curvature is given by $F_{\mu\nu}=\epsilon_{\alpha \beta \gamma} \hat{d}_\alpha \partial_{\mu} \hat{d}_\beta \partial_{\nu} \hat{d}_\gamma$. In contrast, for the $4\times 4$ Dirac Hamiltonian, $\epsilon_{\alpha \beta \gamma} \hat{d}_\alpha \partial_{\mu} \hat{d}_\beta \partial_{\nu} \hat{d}_\gamma$ can always be related to some components of the non-Abelian Berry/Euler curvature. This relation establishes a connection between the non-Abelian quantum metric and the non-Abelian Berry/Euler curvature. Thus, it provides an alternative approach to calculate topological invariants, the Chern number for $CP$ symmetric Hamiltonian and the Euler class for $C_2T$ symmetric Hamiltonian, as discussed in the following sections.

\section{Topological semimetal with $CP$ symmetry}
\label{sec3}
Symmetries such as time reversal ($T$), particle-hole ($C$), twofold rotation ($C_2$), and inversion ($P$) play a fundamental role in topological physics. The combined symmetries of $CP$ and $C_2T$ are of particular importance. In Ref. \cite{YXZhao2016}, the researchers have developed a unified theory to describe the topological properties of nodal structures protected by $CP$ and $C_2T$ symmetries. Let us first focus on a Hamiltonian that is symmetric under $CP$:
\begin{equation}
\begin{aligned}
PH(\textbf{k})P^{-1}&=H(-\textbf{k}),  \\
CH(\textbf{k})^{*}C^{-1}&=-H(-\textbf{k}),    \\
(CP)H(\textbf{k})^{*}(CP)^{-1}&=-H(\textbf{k}).
\end{aligned}\label{CP}
\end{equation}
where $P$ and $C$ are unitary operators. We consider a concrete lattice Hamiltonian
\begin{equation}
\label{hcp}
H_{CP}^{(n)}/\hbar\Omega_0=\frac{\alpha_x}{2}(d_{-}^{n}+d_{+}^{n})\Gamma_{x}+i\frac{\alpha_y}{2}(d_{-}^{n}-d_{+}^{n})\Gamma_{y}+\alpha_z d_z \Gamma_z,
\end{equation}
where $d_{\pm}^n=(d_x \pm id_y)^n$, $\alpha_{x,y,z}=\pm 1$,
$d_x=2t\sin k_x$, $d_y=2t\sin k_y$, and $d_z=2t(M_z-\cos k_x-\cos k_y-\cos k_z$). $M_z$ is a dimensionless and tunable parameter, $t$ is hopping energy, and unless specifically mentioned, we simply set $t=1/2$ for following calculations. $\Gamma_x=\sigma_x \otimes s_x$, $\Gamma_y=\sigma_0 \otimes s_y$, $\Gamma_z=\sigma_x \otimes s_z$. Here $\hbar\Omega_0$ is the irrelevant energy unit, the time unit is given by $2\pi/\Omega_0$, and we set $\hbar=\Omega_0=1$ hereafter. This model Hamiltonian belongs to $2Z$ classification \cite{YXZhao2016} with $CP=\sigma_z \otimes s_0 \mathcal{K}$ and $(CP)^2=1$, where $\mathcal{K}$ is the conjugate operator. The energy spectrum is obtained as
\begin{equation}
E_{\pm}=\pm \sqrt{(d_x^2+d_y^2)^n+d_z^2}.
\end{equation}
This Hamiltonian exhibits multiple-Weyl monopoles at four-fold degenerate points $E_{\pm}=0$, and their topological charges are characterized by the first Chern number. The distribution and separation of these multiple-Weyl points within the first Brillouin zone (FBZ) are controlled by the parameter $M_z$. When $M_z=0$, there are four multiple-Weyl points located at $(\pi,0,\pm \pi/2)$ and $(0,\pi,\pm \pi/2)$. As $M_z$ increases, these monopoles start to move within the FBZ. For $M_z=1$, there exist three monopoles at $(\pi,0,0)$, $(0,\pi,0)$, and $(0,0,\pi)$ with a monopole charge $Q=0$. When $M_z=2$, only two monopoles remain at $(0,0,\pm \pi/2)$ with opposite topological charges. The two monopoles move toward $(0,0,0)$ and eventually combine when $M_z=3$, opening a gap. After that, the system becomes a topologically trivial insulator.

\begin{figure}
\centering
\includegraphics[width=8cm]{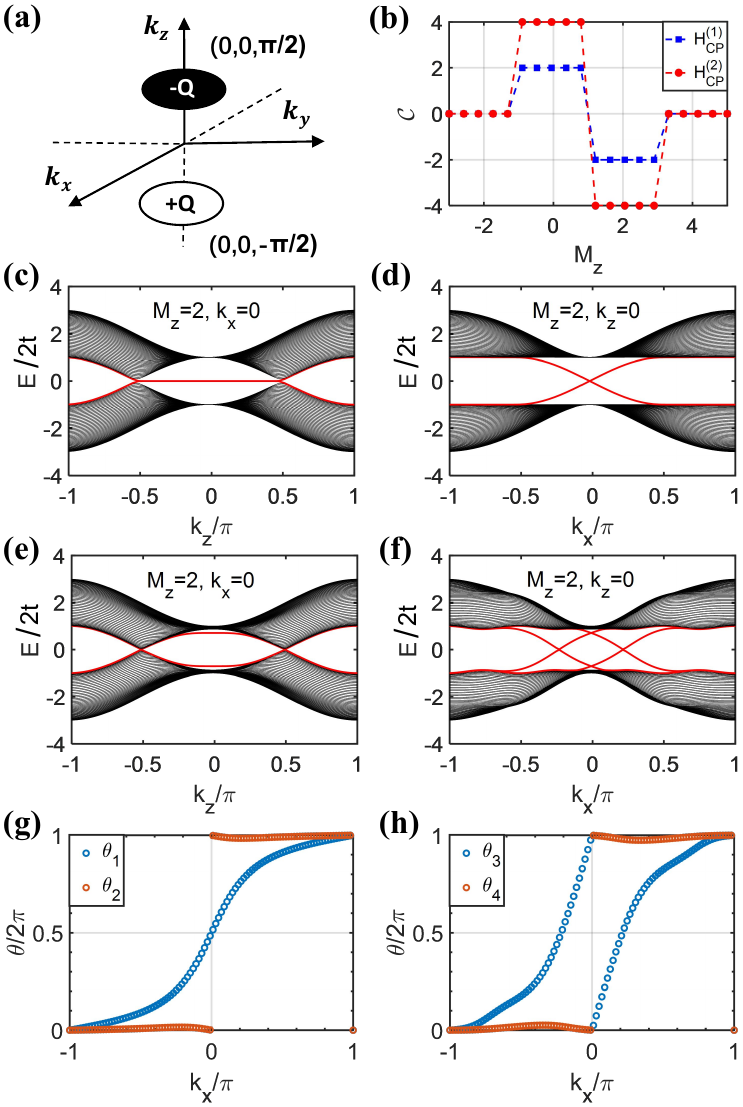}
\caption{(a) The multi-Weyl monopoles of Hamiltonian $H_{CP}^{(n)}$ in FBZ for $M_z=2$, the ellipses stand for monopoles and $\pm Q$ mean their topological charges. $Q=2$ for $H_{CP}^{(1)}$, $Q=4$ for $H_{CP}^{(2)}$. (b) The first Chern number $\mathcal{C}$ for Hamiltonian $H_{CP}^{(1)}$ and $H_{CP}^{(2)}$ against the parameter $M_z$, $k_z=0$. The bulk state and surface state for Hamiltonian $H_{CP}^{(1)}$ and $H_{CP}^{(2)}$. (c) The energy spectra for $H_{CP}^{(1)}$, $M_z=2$, $k_x=0$. (d) for $H_{CP}^{(1)}$, $M_z=2$, $k_z=0$. (e) The bulk state and surface state for Hamiltonian $H_{CP}^{(2)}$, with $M_z=2$, $k_x=0$. (f)  for Hamiltonian $H_{CP}^{(2)}$, with $M_z=2$, $k_z=0$. (g) $\theta(k_x)$, the phases of the eigenvalues of the Wilson loops.  $\theta_1$ and $\theta_2$ are for Hamiltonian $H_{CP}^{(1)}$ in Eq.~(\ref{H_1}) with $M_z=2$ (topological phase) and $M_z=4$ (trivial phase), $k_z=0$.  (h) $\theta_3$ and $\theta_4$ are  for Hamiltonian $H_{CP}^{(2)}$ in Eq.~(\ref{H_2}) with $M_z=2$ (topological phase) and $M_z=4$ (trivial phase), $k_z=0$. $\alpha_x=\alpha_y=\alpha_z=1$ for all panels.}
\label{fig1}
\end{figure}

Considering $M_z=2$, there are two multi-Weyl points located at $\mathbf{K}_{\pm}=(0,0,\pm \frac{\pi}{2})$. The topological charges $Q$ are defined in terms of the Chern number on a sphere $\mathcal{S}^2$ enclosing the multi-Weyl points, as shown in Fig.~\ref{fig1}(a). The low energy effective Hamiltonian near the multi-Weyl points is given by
\begin{equation}
\label{hcpe}
\mathcal{H}_{CP,\pm}^{(n)}=\frac{\alpha_x}{2}(q_{-}^{n}+q_{+}^{n})\Gamma_{x}+i\frac{\alpha_y}{2}(q_{-}^{n}-q_{+}^{n})\Gamma_{y}\pm\alpha_z q_z \Gamma_z,
\end{equation}
where $q_{\pm}^n=(q_x \pm iq_y)^n$, $\textbf{q}_{\pm}=\textbf{k}-\textbf{K}_{\pm}$. The corresponding energy spectrum is $\mathcal{E}_{\pm}=\sqrt{(q_x^2+q_y^2)^n+q_z^2}$.
We can parametrize the momentum space as
\begin{equation}
\label{para}
\begin{aligned}
q_x&=(q\sin \theta)^{\frac{1}{n}}\sin \phi,    \\
q_y&=(q\sin \theta)^{\frac{1}{n}}\cos \phi,    \\
q_z&=q\cos \theta,
\end{aligned}
\end{equation}
where $q=\mathcal{E}_{+}=\sqrt{(q_x^2+q_y^2)^n+q_z^2}$, $\theta\in(0,\pi]$ and $\phi\in(0,2\pi]$ are two spheral angles of a $\mathcal{S}^2$ sphere. For the multi-Weyl points, the topological charge is
\begin{equation}
Q=\frac{1}{2\pi}\int_{\mathcal{S}^2} \text{Tr} (F_{\theta\phi})d\theta d\phi.
\end{equation}
 In Eq.~(\ref{gfre}), we have found that there is a deep connection between non-Abelian quantum metric and unit Bloch vector. For the effective Hamiltonian in Eq. (\ref{hcpe}), $\text{Tr}(F_{\theta\phi})_{\pm}=\pm\epsilon_{\alpha \beta \gamma} \hat{q}_\alpha \partial_{\theta} \hat{q}_\beta \partial_{\phi} \hat{q}_\gamma$ (with $\hat{\mathbf{q}}\equiv \mathbf{q}/|\mathbf{q}|=(\hat{q}_1,\hat{q}_2,\hat{q}_3)$), and $\operatorname{sgn}((\text{Tr} F_{\theta\phi})_{\pm})=\mp 1$. Then Eq.~(\ref{gfre}) can be rewritten as
\begin{equation}
\sqrt{\det (G_{\theta \phi})_{\pm}}=\mp \text{Tr}(F_{\theta\phi})_{\pm},
\end{equation}
 the corresponding topological charges are given by
\begin{equation}
Q_{\pm}=\mp \frac{1}{2\pi}\int \sqrt{\det (G_{\theta\phi})_{\pm}} d\theta d\phi.
\label{g-c}
\end{equation}
The non-Abelian Berry curvature $F_{\theta \phi}$ is a $2\times 2$ matrix,                    $\text{Tr}(F_{\theta\phi})_{\pm}=\mp\sqrt{\det (G_{\theta \phi})_{\pm}}=\mp n\sin \theta\text{sgn}(\alpha_x\alpha_y\alpha_z)$. One has
\begin{equation}
Q_{\pm}=\mp 2n \text{sgn} (\alpha_x \alpha_y \alpha_z),
\end{equation}
which characterizes the $2Z$ nature of this Hamiltonian.
Without loss of generality, we consider $n=1,2$ and $\alpha_x=\alpha_y=\alpha_z=1$ for the Hamiltonian in Eq.~(\ref{hcp}). For $n=1$,
\begin{equation}
\label{H_1}
H_{CP}^{(1)}=d_x\Gamma_x+d_y\Gamma_y+d_z\Gamma_z.
\end{equation}
When $M_z=2$, there are two multi-Weyl points located at $\mathbf{K}_{\pm}=(0,0,\pm \pi/2)$, with topological charges $Q_{\pm}=\mp 2$, as shown in Fig.~\ref{fig1}(a). The effective Hamiltonian near $\mathbf{K}_{\pm}$ is given by
\begin{equation}
\mathcal{H}_{CP,\pm}^{(1)}=q_x\Gamma_x+q_y\Gamma_y\pm q_z\Gamma_z,
\label{Hcpef}
\end{equation}
where $\mathbf{q}_{\pm}=\mathbf{k}-\mathbf{K}_{\pm}$.
We plot the energy spectrum with open boundary along $y$ direction for Hamiltonian $H_{CP}^{(1)}$ in Fig.~\ref{fig1}(c). There present Fermi arcs connecting the monopoles, and the dispersion $\mathcal{E}_{1,\pm}$ near them is linear in $k_x$, $k_y$ and $k_z$ directions
\begin{equation}
\mathcal{E}_{1,\pm}=\pm \sqrt{k_x^2+k_y^2+(k_z-\frac{\pi}{2})^2}.
\end{equation}
By taking a slice of this 3D model, e.g., fixing $k_z=0$, it reduces to a 2D model. For reduced 2D Hamiltonian in Eq.~(\ref{hcp}), the general relation in Eq.~(\ref{gfre}) transformed to
\begin{equation}
\sqrt{\det G_{xy}}=|\text{Tr}(F_{xy})|.
\end{equation}
The topological nature of this 2D four-band Hamiltonian is captured by the Chern number 
\begin{equation}
\begin{aligned}
\mathcal{C}&=\frac{1}{2\pi} \int_{B Z} \text{Tr}(F_{xy}) dk_x dk_y   \\
&=\frac{1}{2\pi} \int_{B Z} \operatorname{sgn}(\text{Tr}(F_{xy}))\sqrt{\det G_{xy}} dk_x dk_y,
\end{aligned}
\label{chern}
\end{equation}
which can be calculated by the non-Abelian QGT. Notably, a similar relation between the Chern number and the Abelian QGT for 2D two-band Chern insulators has been obtained by the same method in Refs. \cite{TOzawa2021,AZhang2022}.

We plot the Chern number against $M_z$ in Fig.~\ref{fig1}(b), the topological phase transitions occur at $M_z=\pm 1, 3$, where there is only one four-fold degenerate point in the FBZ. For $M_z \in (-1,1) \cup (1,3)$, the Chern number is nonzero, indicating Chern insulator phases.
The energy spectrums and boundary states with an open boundary along the $y$ direction for Hamiltonian $H_{CP}^{(1)}$ are shown in Fig.~\ref{fig1}(d). In this case, there are gapless edge states that connect the conduction and valence bands.

For $n=2$ in Eq.~(\ref{hcp}), the Hamiltonian reads
\begin{equation}
\label{H_2}
H_{CP}^{(2)}=(d_x^2-d_y^2)\Gamma_x+2d_xd_y\Gamma_y+d_z\Gamma_z.
\end{equation}
Similar to $H_{CP}^{(1)}$, the emergence and locations of multi-Weyl points for $H_{CP}^{(2)}$ are controlled by the parameter $M_z$. For $M_z=2$, two monopoles locate at $(0,0,\pm\pi/2)$, and the corresponding topological charges are $Q_{\pm}=\mp 4$, as illustrated in Fig.~\ref{fig1}(a).
The effective Hamiltonian near $\mathbf{K}_{\pm}=(0,0,\pm\pi/2)$ reads
\begin{equation}
\mathcal{H}_{CP,\pm}^{(2)}=(q_x^2-q_y^2)\Gamma_x+2q_xq_y\Gamma_y\pm q_z\Gamma_z,
\end{equation}
where $\mathbf{q}_{\pm}=\mathbf{k}-\mathbf{K}_{\pm}$.
The energy spectrum with open boundary along the $y$ direction for Hamiltonian $H_{CP}^{(2)}$ is shown in Fig.~\ref{fig1}(e). There also emerges Fermi arcs connecting the multi-Weyl points, but the dispersions $\mathcal{E}_{2,\pm}$ near them are quadratic along $k_{x,y}$ and linear along $k_z$, that is
\begin{equation}
\mathcal{E}_{2,\pm}=\pm \sqrt{(k_x^2+k_y^2)^2+(k_z-\frac{\pi}{2})^2}.
\end{equation}
For fixed $k_z=0$, a reduced 2D model can be derived. We show the first Chern number of this 2D Hamiltonian in Fig.~\ref{fig1}(b), and it can also be calculated from Eq.~(\ref{chern}). The Chern number of $H_{CP}^{(2)}$ is twice as much as that of $H_{CP}^{(1)}$. In Fig.~\ref{fig1}(f), we show the bulk and edge states for the topological insulator phase with symbolic gapless boundary modes crossing the energy gap. Alternatively, we can work out the transition function in terms of the Wilson loop for 2D gapped subsystems.
\begin{equation}
W\left(k_x\right)=\mathcal{P} \exp \int_{-\pi}^\pi d k_y \mathcal{A}_y\left(k_x, k_y\right),
\end{equation}
where the $\mathcal{P}$ indicates the path order.
In Fig.~\ref{fig1}(g-h), we numerically plot the phases of the eigenvalues of the Wilson loop. The results show that the winding numbers equal half of the corresponding Chern numbers.

\section{Topological semimetal with combined $C_2 T$ symmetry}
\label{sec4}
\begin{figure}
\centering
\includegraphics[width=8cm]{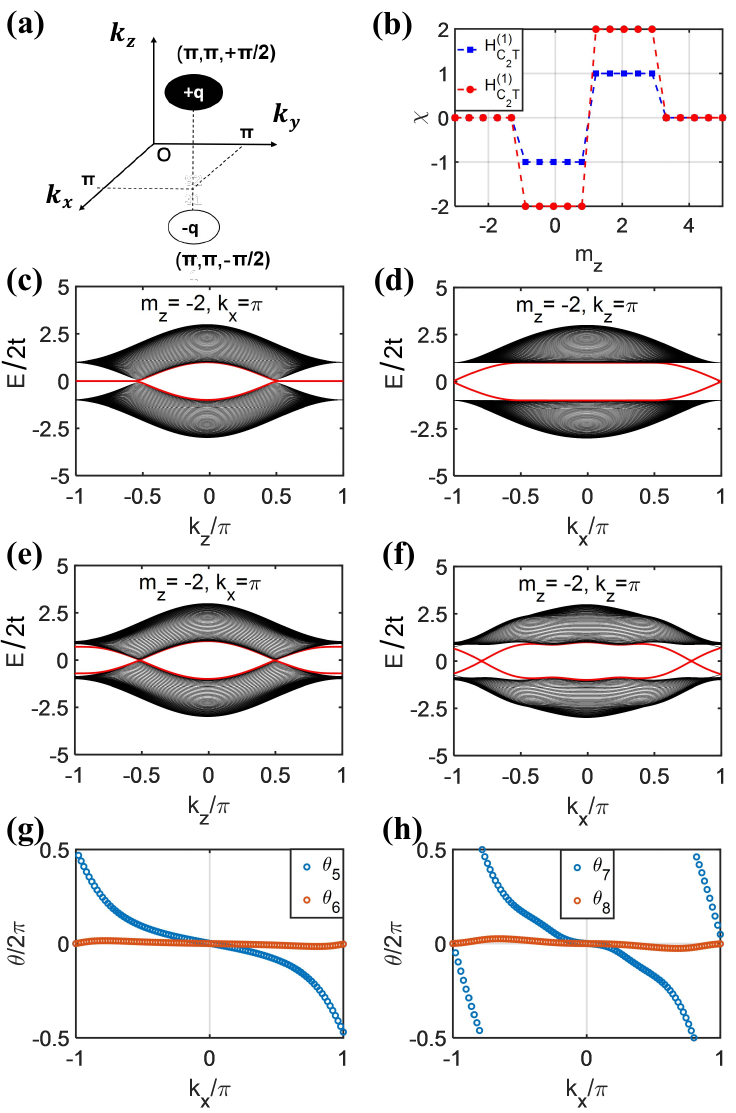}
\caption{(a) The monopoles of Hamiltonian $H_{C_2T}^{(n)}$ in FBZ for $m_z=-2$, the ellipses stand for monopoles and $\pm q$ mean their topological charges, here $q=n$ for $H_{C_2T}^{(n)}$ (b) The Euler class for Hamiltonian $H_{C_2T}^{(1)}$ and $H_{C_2T}^{(2)}$, $k_z=0$. (c) The energy spectra for Hamiltonian $H_{C_2T}^{(1)}$, with $m_z=-2$, $k_x=\pi$, it is topological Euler semimetal. (d)  for Hamiltonian $H_{C_2T}^{(1)}$, with $m_z=-2$, $k_z=\pi$. It is Euler insulator. (e) The bulk state and surface state for Hamiltonian $H_{C_2T}^{(2)}$, with $m_z=-2$, $k_x=\pi$. (f)  for Hamiltonian $H_{C_2T}^{(2)}$, with $m_z=-2$, $k_z=\pi$. (g) $\theta_5$ and $\theta_6$ are transition functions for Hamiltonian $H_{C_2T}^{(1)}$ in Eq.~(\ref{hct}) with $m_z=2$ (topological phase) and $m_z=4$ (trivial phase), $k_z=0$. (h) $\theta_7$ and $\theta_8$ are for Hamiltonian $H_{C_2T}^{(2)}$ in Eq.~(\ref{hct}) with $m_z=2$ (topological phase) and $m_z=4$ (trivial phase), $k_z=0$.  $\alpha_x=\alpha_y=\alpha_z=1$ for all panels. }
\label{fig2}
\end{figure}

The 3D topological Euler semimetals and 2D Euler insulators are novel kinds of topological phases. Euler insulators host multi-gap topological phases, which are quantified by a quantized Euler class in their bulk and have been recently experimentally realized in synthetic systems \cite{BJiang2021,WZhao2022,BJiang2022}. The $C_{2}T$ symmetry implies the existence of a basis in which the $C_{2}T$ symmetric Bloch Hamiltonian is a real matrix. We consider a concrete $C_2T$ symmetric Hamiltonian
\begin{equation}
\label{hct}
H_{C_{2}T}^{(n)}/\hbar\Omega_0=\frac{\alpha_x}{2}(d_{-}^{n}+d_{+}^{n})\tilde{\Gamma}_{x}+i\frac{\alpha_y}{2}(d_{-}^{n}-d_{+}^{n})\tilde{\Gamma}_{y}+\alpha_z d_z \tilde{\Gamma}_z,
\end{equation}
where $\tilde{\Gamma}_x=\sigma_0 \otimes s_z$, $\tilde{\Gamma}_y=\sigma_y \otimes s_y$, and $\tilde{\Gamma}_z=\sigma_0 \otimes s_x$. The energy spectrum $E_{\pm}=\pm \sqrt{(d_x^2+d_y^2)^n+d_z^2}$. Without loss of generality, we choose $C_{2}T=\hat{\mathcal{K}}$.
Under $C_2T$ symmetry, the Hamiltonian and the Bloch wave functions can always be constrained to real-valued by a unitary transformation, which makes the first Chern number of it equals to zero.
For $C_{2}T$ symmetric Hamiltonian, the Euler class has been defined to characterize topological phase transitions. If $\left|\psi_{1}(\boldsymbol\lambda)\right\rangle$ and $\left|\psi_{2}(\boldsymbol\lambda)\right\rangle$ are real Bloch states of a pair of global degenerate energy bands, their Euler curvature (also called Euler form) \cite{Unal2020,ABouhon2020,MEzawa20211} is given by
\begin{equation}
\operatorname{Eu}(\boldsymbol\lambda)=\left\langle\nabla \psi_1 (\boldsymbol\lambda)|\times| \nabla \psi_2 (\boldsymbol\lambda)\right\rangle.
\end{equation}
When the base manifold is 2D and parameterized by $\boldsymbol\lambda=\left\{\lambda_{\mu},\lambda_{\nu}\right\}$, the Euler curvature is
\begin{equation}
\label{Eu1}
\operatorname{Eu}(\boldsymbol\lambda)=\left\langle\partial_{\mu} \psi_1(\boldsymbol\lambda) \mid \partial_{\nu} \psi_2(\boldsymbol\lambda)\right\rangle-\left\langle\partial_{\nu} \psi_1(\boldsymbol\lambda) \mid \partial_{\mu} \psi_2(\boldsymbol\lambda)\right\rangle.
\end{equation}
The integral of the Euler curvature over a 2D manifold defines the integer topological invariant, which is called the Euler class
\begin{equation}
\label{Eulam}
\chi=\frac{1}{2\pi} \int \operatorname{Eu}(\boldsymbol\lambda) d\lambda_{\mu}d\lambda_{\nu}.
\end{equation}
Consider the corresponding dimensionless parameter $m_z=-2$, it is a topological Euler semimetal phase with the monopole charges characterized by the Euler class. There are two monopoles located at $\tilde{\mathbf{K}}_{\pm}=(\pi,\pi,\pm \pi/2)$. The topological charges $q$ are defined in terms of the Euler class $\chi$ on a sphere $\mathcal{S}^2$ enclosing the monopoles, as shown in Fig.~\ref{fig2}(a). The low-energy effective Hamiltonians near the monopoles are given by
\begin{equation}
\mathcal{H}_{C_2T,\pm}^{(n)}=-\frac{\alpha_x}{2}(\tilde{q}_{-}^{n}+\tilde{q}_{+}^{n})\tilde{\Gamma}_{x}-i\frac{\alpha_y}{2}(\tilde{q}_{-}^{n}-\tilde{q}_{+}^{n})\tilde{\Gamma}_{y}\pm\alpha_z \tilde{q}_z \tilde{\Gamma}_z,
\label{c2teff}
\end{equation}
where $\tilde{q}_{\pm}^n=(\tilde{q}_x \pm i\tilde{q}_y)^n$, $\tilde{\textbf{q}}_{\pm}=\textbf{k}-\tilde{\textbf{K}}_{\pm}$. The energy spectrum $\tilde{\mathcal{E}}_{\pm}=\pm\sqrt{(\tilde{q}_x^2+\tilde{q}_y^2)^n+\tilde{q}_z^2}$.
By parametrizing the momentum space with spherical coordinates as in Eq.~(\ref{para}),
with $\boldsymbol\lambda=\left\{\theta,\phi\right\}$ in Eq.~(\ref{Eulam}),
the monopole charge is described by Euler class
\begin{equation}
q=\frac{1}{2\pi}\int_{\mathcal{S}^2} \mathrm{Eu}(\theta, \phi) d\theta d\phi.
\end{equation}
Here $\operatorname{Eu}(\theta, \phi)$ is the Euler curvature in spherical coordinates
\begin{equation}
\operatorname{Eu}(\theta, \phi)=\left\langle\partial_{\theta} \psi_1 \mid \partial_{\phi} \psi_2\right\rangle-\left\langle\partial_{\phi} \psi_1 \mid \partial_{\theta} \psi_2\right\rangle,
\end{equation}
where $\left|\psi_{1}\right\rangle$ and $\left|\psi_{2}\right\rangle$ are the degenerate ground states of Hamiltonian.
The Euler curvature for $\mathcal{H}_{C_2T,\pm}^{(n)}$
\begin{equation}
\operatorname{Eu}_{\pm}=\mp\frac{1}{2}\epsilon_{\alpha \beta \gamma} \hat{\tilde{q}}_\alpha \partial_{\theta} \hat{\tilde{q}}_\beta \partial_{\phi} \hat{\tilde{q}}_\gamma=\pm\frac{i}{2}(F_{\theta\phi}^{12}-F_{\theta\phi}^{21}),
\end{equation}
where $\hat{\tilde{\mathbf{q}}}\equiv \tilde{\mathbf{q}}/|\tilde{\mathbf{q}}|=(\hat{\tilde{q}}_1,\hat{\tilde{q}}_2,\hat{\tilde{q}}_3)$.
From Eq.~(\ref{gfre}), we obtain the relation between non-Abelian quantum metric and Euler curvature as
\begin{equation}
\sqrt{\det (G_{\theta\phi})_{\pm}}=|\epsilon_{\alpha \beta \gamma} \hat{d}_\alpha \partial_{\theta} \hat{d}_\beta \partial_{\phi} \hat{d}_\gamma|=\pm2\operatorname{Eu}_{\pm}.
\end{equation}
Thus the topological charges can also be extracted from non-Abelian quantum metric as
\begin{equation}
q_{\pm}=\pm \frac{1}{4\pi}\int \sqrt{\det (G_{\theta\phi})_{\pm}} d\theta d\phi.
\end{equation}
From the Euler curvature $\operatorname{Eu}_{\pm}=\pm\sqrt{\det (G_{\theta\phi})_{\pm}}/2=\pm n\sin \theta /2$, we have
\begin{equation}
q_{\pm}=\pm n\text{sgn}(\alpha_x\alpha_y\alpha_z),
\end{equation}
as shown in Fig.~\ref{fig2}(a).
Consider $n=1,2$ and $\alpha_x=\alpha_y=\alpha_z=1$, we have the Hamiltonians
\begin{equation}
\label{hct}
\begin{aligned}
H_{C_2T}^{(1)}&=d_x\tilde{\Gamma}_x+d_y\tilde{\Gamma}_y+d_z\tilde{\Gamma}_z,   \\
H_{C_2T}^{(2)}&=(d_x^2-d_y^2)\tilde{\Gamma}_x+2d_xd_y\tilde{\Gamma}_y+d_z\tilde{\Gamma}_z.
\end{aligned}
\end{equation}
The effective Hamiltonians near $\tilde{\mathbf{K}}_{\pm}$ are given by
\begin{equation}
\begin{aligned}
\mathcal{H}_{C_2T,\pm}^{(1)}&=-\tilde{q}_x\tilde{\Gamma}_x-\tilde{q}_y\tilde{\Gamma}_y\pm \tilde{q}_z\tilde{\Gamma}_z, \\
\mathcal{H}_{C_2T,\pm}^{(2)}&=-(\tilde{q}_x^2-\tilde{q}_y^2)\tilde{\Gamma}_x-2\tilde{q}_x\tilde{q}_y\tilde{\Gamma}_y \pm \tilde{q}_z\tilde{\Gamma}_z,
\end{aligned}
\label{c2teff}
\end{equation}
where $\tilde{\mathbf{q}}_{\pm}=\mathbf{k}-\tilde{\mathbf{K}}_{\pm}$.
In Figs.~\ref{fig2} (c-d), we plot the energy spectrum of $H_{C_2T}^{(1)}$ and $H_{C_2T}^{(2)}$, with open boundary condition along the $y$ direction. The Fermi arcs connect the energy degenerate points. They both are topological Euler semimetal phases.
Taking a slice with fixed $k_z=0$ for a reduced 2D model, one has the Euler curvature
\begin{equation}
\operatorname{Eu}(k_x,k_y)=\frac{1}{2}\epsilon_{\alpha \beta \gamma} \hat{d}_\alpha \partial_{x} \hat{d}_\beta \partial_{y} \hat{d}_\gamma=-\frac{i}{2}(F_{xy}^{12}-F_{yx}^{21}),
\end{equation}
where $\sqrt{\operatorname{det} G_{xy}}=|\epsilon_{\alpha \beta \gamma} \hat{d}_\alpha \partial_{x} \hat{d}_\beta \partial_{y} \hat{d}_\gamma|$.
The corresponding Euler class reads
\begin{equation}
\begin{aligned}
\chi & =\frac{1}{2 \pi} \int_{BZ} \operatorname{Eu} dk_{x} dk_{y}
 =\frac{-i}{4 \pi} \int_{BZ} (F_{xy}^{12}-F_{xy}^{21}) dk_{x} dk_{y} \\
& =\frac{1}{4\pi} \int_{BZ} \operatorname{sgn}\left(\operatorname{Im}\left(F_{xy}^{12}-F_{xy}^{21}\right)\right) \sqrt{\operatorname{det} G_{xy}} dk_{x} dk_{y}.
\end{aligned}
\end{equation}
Figure~\ref{fig2}(b) shows the relation between the parameter $m_z$ and Euler class $\chi_{1,2}$ for $H_{C_2T}^{(1,2)}$ with $\chi_2 =2\chi_1$. The topological phase transitions occur when the energy band gap close, e.g., $m_z=\pm1,3$. For $m_z \in (-1,1) \cup (1,3)$, Euler class $\chi_1$ and $\chi_2$ are nonzero, they are topological Euler insulator phases.
Figures~\ref{fig2}(e-f) show the numerical results for $H_{C_2T}^{(1)}$ and $H_{C_2T}^{(2)}$ with fixed $m_z=2$ and $k_z=0$.
We exhibit the bulk states and boundary states of the two Hamiltonians, the corresponding topologically metallic edge states emerge.
We can also calculate the transition function of the real bundles for 2D Euler insulator through the Wilson loops
\begin{equation}
W\left(k_x\right)=\mathcal{P} \exp \int_{-\pi}^\pi d k_y \mathcal{A}^{R}_y\left(k_x, k_y\right),
\end{equation}
where $\mathcal{A}^{R}_y$ is a real Berry connection, and the component $(\mathcal{A}_{y}^R)_{ij}=\left\langle\psi_i \mid \partial_{y} \psi_j\right\rangle$, with $\left|\psi_{i}\right\rangle$ and $\left|\psi_{j}\right\rangle$ being the degenerate ground states. The transition function $\theta(k_x)$ as a function of $k_x$ can be extracted from the Wilson loops $e^{-i\sigma_2 \theta(k_x)}=W(k_x)$. The numerical results of transition functions are illustrated in Fig.~\ref{fig2}(g-h).  It can be observed that the winding numbers of the transition functions are equal to the corresponding Euler classes.

\section{Scheme to extract non-Abelian quantum geometric tensor}
\label{sec5}

We first consider the non-adiabatic response related to the non-Abelian QGT. It has been shown that its imaginary part is linked to the so-called generalized force \cite{Abigail2018,Kolodrubetz2016}. Below we show that the real part of the non-Abelian QGT is related to the energy fluctuation. Thus, these non-adiabatic responses provide a new method to extract all the components of the non-Abelian QGT.

Consider a Hamiltonian $H(\boldsymbol\lambda)$ parameterized by $\boldsymbol\lambda$. The adiabatic perturbation theory regards the quantum adiabatic approximation as the zeroth-order case and describes a perturbation expansion in terms of the small changing velocity of the parameter $\boldsymbol\lambda$. This theory has been generalized to the case of degenerate ground states \cite{Abigail2018,Kolodrubetz2016}.
If start at one of the degenerate ground state and ramp the parameter $\boldsymbol\lambda$ slowly with time, $\dot{\boldsymbol\lambda}\approx 0$.  Consider a path such that an adiabatic traversal would yield a particular ground state $\left|\psi_{i}(\boldsymbol\lambda(0))\right\rangle$ \cite{Kolodrubetz2016}.  Tracing the same path at a finite rate, the ground state component of the wave function remains unchanged up to order $\dot{\boldsymbol\lambda}$. At time $t$, $\boldsymbol\lambda=\boldsymbol\lambda(t)$, and the quantum state can be written as \cite{Kolodrubetz2016}:
\begin{equation}
\label{A}
|\psi(\boldsymbol\lambda)\rangle \approx\left|\psi_{i}(\boldsymbol\lambda)\right\rangle+i \dot{\lambda}_{\mu} \sum_{n \neq i}\left|\psi_{n}(\boldsymbol\lambda)\right\rangle \frac{\left\langle\psi_{n}(\boldsymbol\lambda) \mid \partial_{\mu} \psi_{i}(\boldsymbol\lambda)\right\rangle}{E_{n}-E_{0}}.
\end{equation}
We can always represent observables as generalized force operators conjugate to some other  coupling $\lambda_{\nu}$: $\mathcal{M}_{\nu}=\partial_{\nu} H(\boldsymbol\lambda)$. At time $t$, the expectation value of $\mathcal{M}_{\nu}$ \cite{Kolodrubetz2016}
\begin{equation}
\label{gen}
\begin{aligned}
\mathcal{M}_{\nu} \equiv\left\langle\psi(\boldsymbol\lambda)\left|\mathcal{M}_{\nu}\right| \psi(\boldsymbol\lambda)\right\rangle\approx\mathcal{M}_{\nu}^{0}+\dot{\lambda}_{\mu} F_{\mu \nu}^{ii},
\end{aligned}
\end{equation}
with $\mathcal{M}_{\nu}^{0}=\left\langle\psi_i(\boldsymbol\lambda)\left|\mathcal{M}_{\nu}\right| \psi_i(\boldsymbol\lambda)\right\rangle$ and
\begin{equation}
F_{\mu \nu}^{ii}=i\sum_{n \in \text{unocc}} \frac{\left\langle \psi_i\left|\partial_{\mu} {H}\right| \psi_n\right\rangle \left  \langle \psi_n\left|\partial_{\nu} {H}\right| \psi_i\right\rangle - (\mu \leftrightarrow \nu )}{\left(E_{n}-E_{0}\right)^{2}}.
\end{equation}
This relation shows that the leading non-adiabatic correction to the generalized force comes from the product of the non-Abelian Berry curvature and the rate of change of the parameter $\boldsymbol\lambda$ \cite{Michael2017}.

For the non-Abelian quantum metric, we find a related observable as the energy fluctuation
\begin{equation}
\label{En}
\begin{aligned}
&\Delta E^{2}=\left\langle\psi(\boldsymbol\lambda)\left|H(\boldsymbol\lambda)^2\right| \psi(\boldsymbol\lambda)\right\rangle-\left\langle\psi(\boldsymbol\lambda)\left|H(\boldsymbol\lambda)\right| \psi(\boldsymbol\lambda)\right\rangle^2\\
&\approx \sum_{\alpha \beta} \dot{\lambda}_{\alpha} \dot{\lambda}_{\beta}\left[\sum_{n\in \text{unocc}}\frac{\left\langle \psi_i\left|\partial_{\alpha} {H}\right| \psi_n\right\rangle\left\langle \psi_n\left|\partial_{\beta}{H}\right| \psi_i\right\rangle}{\left(E_{n}-E_{0}\right)^{2}}\right]  \\
&=\sum_{\alpha \beta}g_{\alpha \beta}^{ii} \dot{\lambda}_{\alpha} \dot{\lambda}_{\beta}.
\end{aligned}
\end{equation}
Namely, the non-Abelian quantum metric defines the leading non-adiabatic correction to the energy fluctuation. In the adiabatic evolution, the energy fluctuation is zero when the system has a well-defined energy. This result is not limited to degenerate ground states and applies to any initial eigenstates \cite{Polkovnikov2012,Kolodrubetz2016,Abigail2018,Kolodrubetz2013,Michael2017}.

The measurement of QGT can be  observed experimentally in optical lattices using ultracold atoms through Bloch state tomography \cite{CRYI2023}. We note that ramps are routinely achieved in ultracold atom systems,  allowing us to set the parameter $\boldsymbol\lambda=\boldsymbol k=(k_x, k_y)$ in the following numerical calculations.
For simplicity, we consider the CP symmetric lattice Hamiltonian $H_{CP}^{(1)}$ and the $C_{2}T$ symmetric Hamiltonian $H_{C_2T}^{(1)}$ with $M_z=m_z=2$ and $k_z=0$ for 2D topological insulator phases.
We first consider the non-Abelian Berry curvature. From definition, the non-Abelian Berry curvature is a $4\times 4$ matrix
\begin{equation}
\begin{aligned}
F
&=\left(\begin{array}{cc}
F_{xx} & F_{xy}
\vspace{1ex}\\
F_{yx} & F_{yy}
\end{array}\right) =\left(\begin{array}{cccc}
F_{xx}^{11} & F_{xx}^{12} & F_{xy}^{11} & F_{xy}^{12}
\vspace{1ex}\\
F_{xx}^{21} & F_{xx}^{22} & F_{xy}^{21} & F_{xy}^{22}
\vspace{1ex}\\
F_{yx}^{11} & F_{yx}^{12} & F_{yy}^{11} & F_{yy}^{12}
\vspace{1ex}\\
F_{yx}^{21} & F_{yx}^{22} & F_{yy}^{21} & F_{yy}^{22}
\end{array}\right).
\end{aligned}
\end{equation}
$F_{xx}=F_{yy}=O_{2\times 2}$, $F_{xy}=-F_{yx}$ and $F_{xy}^{12}=(F_{xy}^{21})^{*}$. Then the non-Abelian Berry curvature simplified to
\begin{equation}
\begin{aligned}
F
=\left(\begin{array}{cccc}
0 & 0 & F_{xy}^{11} & F_{xy}^{12}
\vspace{1ex}\\
0 & 0 & (F_{xy}^{12})^{*} & F_{xy}^{22}
\vspace{1ex}\\
-F_{xy}^{11} & -F_{xy}^{12} & 0 & 0
\vspace{1ex}\\
-(F_{xy}^{12})^{*} & -F_{xy}^{22} & 0 & 0
\end{array}\right).
\end{aligned}
\end{equation}

\begin{figure}
\centering
\includegraphics[width=0.48\textwidth]{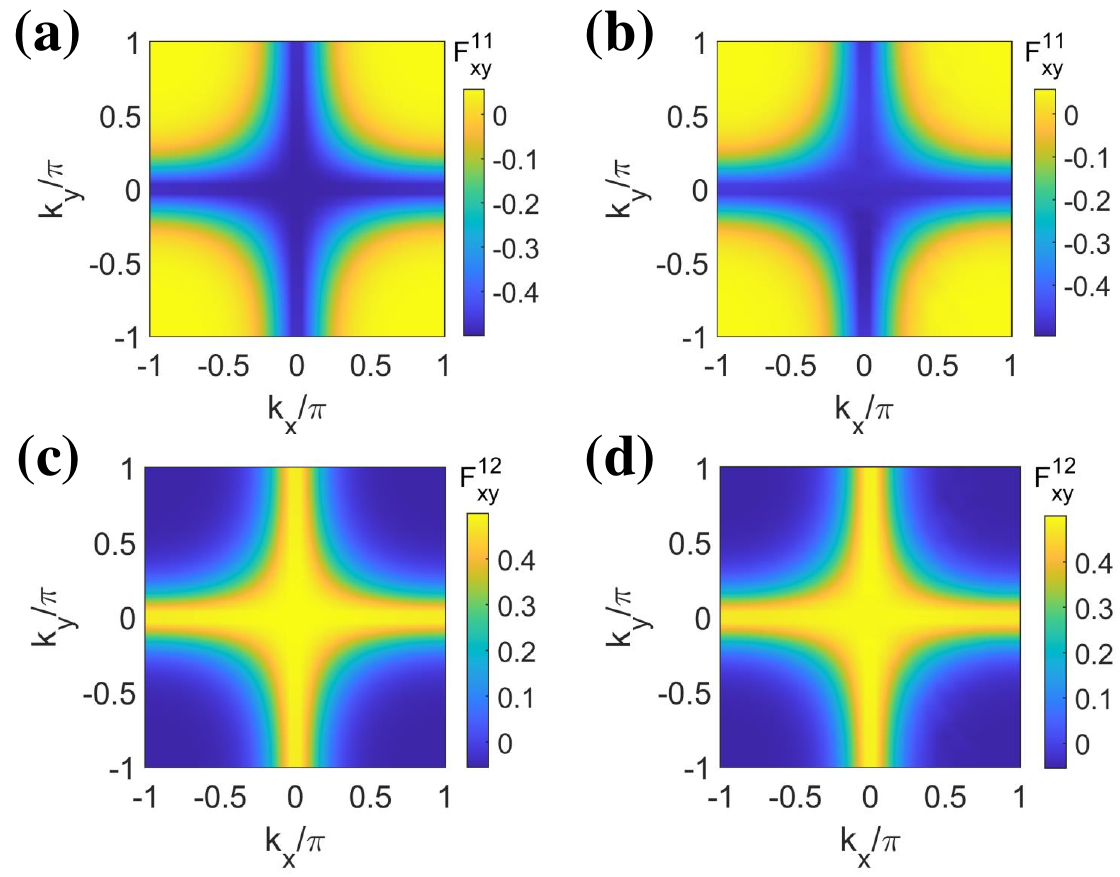}
\caption{(a) The analytical result of $F_{xy}^{11}$ for Hamiltonian $H_{CP}^{(1)}$, $\Im(F_{xy}^{11})=0$. (b) The numerical result of $F_{xy}^{11}$ for Hamiltonian $H_{CP}^{(1)}$. (c) The analytical result of $\Im (F_{xy}^{12})$ for Hamiltonian $H_{C_2T}^{(1)}$. (d) The numerical  result of $\Im(F_{xy}^{12})$ for Hamiltonian $H_{C_2T}^{(1)}$. The numerical results are obtained from full-time-dynamics simulations, and $M_z=m_z=2$, $k_z=0$ for all panels.}
\label{fig3}
\end{figure}

Now we briefly show how to measure the non-Abelian Berry curvature. Assume the initial state at $\left|\psi_{i}(\boldsymbol k(0))\right\rangle$ and ramp the parameter $\boldsymbol k$ along $k_{\mu}$ direction as follows: $k_{\mu}(t)=k_{\mu}(0)+\frac{v^2t^2}{2\pi}$, the ramping velocity is $\dot{k}_{\mu}(t)=\frac{v^2t}{\pi}$. The initial state will evolves with the time-dependent Hamiltonian until the final time $t_{f}=\frac{\pi}{v}$ (in units of $2\pi/\Omega_0$), with the final velocity $\dot{k}_{\mu}(t_f)=v$. From Eq.~\eqref{gen}, we can directly get
\begin{equation}
\label{Fii}
F_{\mu \nu}^{ii}=(\mathcal{M}_{\nu}-\mathcal{M}_{\nu}^0)/v.
\end{equation}
For the component $F_{\mu \nu}^{ij}$, it has following relation \cite{Kolodrubetz2016}
\begin{equation}
F_{\mu \nu}^{ij}=\frac{2 i F_{\mu \nu}^{mm}+2 F_{\mu \nu}^{nn}-(1+i)\left(F_{\mu \nu}^{ii}+F_{\mu \nu}^{jj}\right)}{2 i},
\end{equation}
with $\left|\psi_{m}(\boldsymbol k)\right\rangle=(\left|\psi_{1}(\boldsymbol k)\right\rangle+\left|\psi_{2}(\boldsymbol k)\right\rangle)/\sqrt{2}$ and  $\left|\psi_{n}(\boldsymbol k)\right\rangle=(\left|\psi_{1}(\boldsymbol k)\right\rangle+i\left|\psi_{2}(\boldsymbol k)\right\rangle)/\sqrt{2}$. $F_{\mu \nu}^{mm}$ and $F_{\mu \nu}^{nn}$ can be extracted from Eq.~\eqref{Fii}, we only should prepare the initial state at $\left|\psi_{m}(\boldsymbol k(0))\right\rangle$ and $\left|\psi_{n}(\boldsymbol k(0))\right\rangle$ respectively. So $F_{\mu \nu}^{ij}$ can be extracted. That is to say, all the components of non-Abelian Berry curvature can be detected from the non-adiabatic effects.
In Fig.~\ref{fig3}, we show some numerical results compared to analytical results. Here we set $k_x(t)=k_x(0)+\frac{v^2t^2}{2\pi}$, $v=0.1$, $t_f=\frac{\pi}{v}$. This kind of ramp can make the initial ramping velocity vanish, and initial evolution is adiabatic, which will lift the oscillation caused by the initial state, as discussed in Ref. \cite{XZhang2023,MDSchroer2014}.   For $H_{CP}^{(1)}$, $F_{xy}^{11}\neq 0$, $F_{xy}^{22}\neq 0$ and $F_{xy}^{12}=F_{xy}^{21}=0$. While for  $H_{C_2T}^{(1)}$, $F_{xy}^{12}\neq 0$, $F_{xy}^{21}\neq 0$  and  $F_{xy}^{11}=F_{xy}^{22}= 0$. So we show the numerical result of $F_{xy}^{11}$ for the Hamiltonian $H_{CP}^{(1)}$, and the numerical result of $F_{xy}^{12}$ for $H_{C_2T}^{(1)}$ in Fig.~\ref{fig3}. For comparison, we also show analytical results. These two results agree well with each other.

\begin{figure}
\centering
\includegraphics[width=0.48\textwidth]{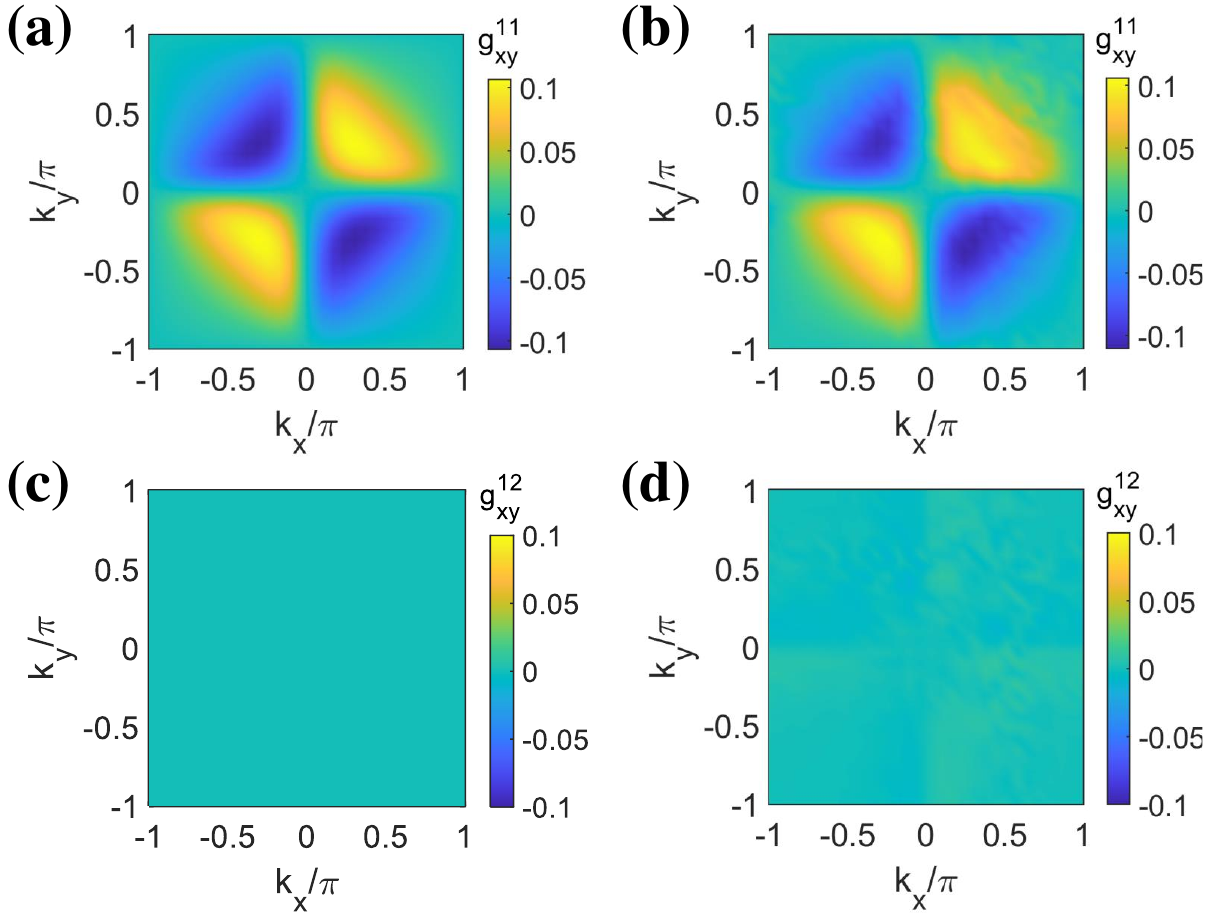}
\caption{(a) The analytical result of $g_{xy}^{11}$ for Hamiltonian $H_{CP}^{(1)}$. (b) The numerical result of $g_{xy}^{11}$ for Hamiltonian $H_{CP}^{(1)}$. (c) The analytical result of $g_{xy}^{12}$ for Hamiltonian $H_{C_2T}^{(1)}$, it is approaching to 0 over the entire BZ. (d) The numerical result of $g_{xy}^{12}$ for Hamiltonian $H_{C_2T}^{(1)}$. The numerical results are obtained from full-time-dynamics simulations, with $M_z=m_z=2$, $k_z=0$. }
\label{fig4}
\end{figure}

We further consider to dynamically extract the non-Abelian quantum metric
\begin{equation}
g =\left(\begin{array}{cc}
g_{xx} & g_{xy}
\vspace{1ex}\\
g_{yx} & g_{yy}
\end{array}\right) =\left(\begin{array}{cccc}
g_{xx}^{11} & g_{xx}^{12} & g_{xy}^{11} & g_{xy}^{12}
\vspace{1ex}\\
g_{xx}^{21} & g_{xx}^{22} & g_{xy}^{21} & g_{xy}^{22}
\vspace{1ex}\\
g_{yx}^{11} & g_{yx}^{12} & g_{yy}^{11} & g_{yy}^{12}
\vspace{1ex}\\
g_{yx}^{21} & g_{yx}^{22} & g_{yy}^{21} & g_{yy}^{22}
\end{array}\right).
\end{equation}
From definition of non-Abelian quantum metric, $g_{xy}=g_{yx}$.
Firstly the initial state is prepared at $\left|\psi_{i}(\boldsymbol k(0))\right\rangle$ and the parameter $\boldsymbol k$ is driven along direction $\lambda_{\mu}$ as follows: $k_{\mu}(t)=k_{\mu}(0)+\frac{v^2t^2}{2\pi}$, and at the final time $t_{f}=\frac{\pi}{v}$, the ramping velocity equals to $v$. From the relation in Eq.~\eqref{En}
\begin{equation}
g_{\mu \mu}^{ii}=\Delta E^2/v^2.
\end{equation}
The energy fluctuation in the final instantaneous state can be measured through fluorescence detection
during optical excitation \cite{XZhang2023}. For example, in order to measure the energy fluctuation, we can get the population of the instantaneous Hamiltonian in repeated experiments, then energy fluctuation can be measured, and so the non-Abelian quantum metric. To extract $g_{\mu \nu}^{ii}$, the initial state is prepared at $\left|\psi_{i}(\boldsymbol k(0))\right\rangle$ and the parameter ramps along $k_{\mu}$ and $k_{\nu}$ directions simutaneously until $t_f=\frac{\pi}{v}$,
\begin{equation}
\begin{aligned}
k_{\mu}(t)&=k_{\mu}(0)+\frac{v^2t^2}{2\pi},   \\
k_{\nu}(t)&=k_{\nu}(0)+\frac{v^2t^2}{2\pi}.
\end{aligned}
\end{equation}
Then $g_{\mu \nu}^{ii}$ is obtained as
\begin{equation}
g_{\mu \nu}^{ii}=(\Delta E^2-g_{\mu \mu}^{ii}v^2-g_{\nu \nu}^{ii}v^2)/2v^2.
\end{equation}
From the definition, we have derived following relation
\begin{equation}
g_{\mu \mu}^{i j}=\frac{2 i g_{\mu \mu}^{mm}+2 g_{\mu \mu}^{nn}-(1+i)\left(g_{\mu \mu}^{i i}+g_{\mu \mu}^{j j}\right)}{2 i}.
\end{equation}
So using above scheme, we can extract $g_{\mu \mu}^{ij}$ from $g_{\mu \mu}^{mm}$, $g_{\mu \mu}^{nn}$, $g_{\mu \mu}^{ii}$ and $g_{\mu \mu}^{jj}$. For $g_{\mu \nu}^{ij}$,
\begin{equation}
g_{\mu \nu}^{i j}=\frac{2 i g_{\mu \nu}^{mm}+2 g_{\mu \nu}^{nn}-(1+i)\left(g_{\mu \nu}^{i i}+g_{\mu \nu}^{j j}\right)}{2 i}.
\end{equation}
If we have already extracted $g_{\mu \nu}^{ii}$, $g_{\mu \nu}^{jj}$, $g_{\mu \nu}^{mm}$ and $g_{\mu \nu}^{nn}$ by using the method mentioned above, then we can extract $g_{\mu \nu}^{ij}$. So all the components of non-Abelian quantum metric can be extracted in experiments by this method. Some numerical results have also been presented in Fig.~\ref{fig4}. Here we consider quadratic ramps \cite{WMa2018} with $k_x(t)=k_x(0)+\frac{v^2t^2}{2\pi}$, $k_y(t)=k_y(0)+\frac{v^2t^2}{2\pi}$, $v=0.1$, $t_f=\frac{\pi}{v}$.  For $H_{CP}^{(1)}$ and $H_{C_2T}^{(1)}$,  $g_{xy}^{11}\neq0$ and $g_{xy}^{12}=0$, here we only show $g_{xy}^{11}$ for $H_{CP}^{(1)}$ and $g_{xy}^{12}$ for $H_{C_2T}^{(1)}$.  The numerical results coincides with analytical results.

The discussions above can be naturally extended to low-energy effective Hamiltonians, and we take $\mathcal{H}_{CP,+}^{(1)}$ in Eq.~(\ref{Hcpef}) and $\mathcal{H}_{C_2T,+}^{(1)}$ in Eq.~(\ref{c2teff}) for examples. In these cases, parameter $\boldsymbol\lambda=(\theta, \phi)$. Set $v=0.1$, final time $t_f=\frac{\pi}{v}$. With the numerically extracted non-Abelian Berry curvature, we obtain the Chern number $\mathcal{C}_{+}=-1.9952 \approx -2$, the Euler class $\chi_{+}=1.0364\approx 1$. Alternatively, these two topological invariants can also be extracted from the non-Abelian quantum metric through the measurement of energy fluctuation. In these two cases, we set $\theta(t)=\theta(0)+\frac{v^2t^2}{2\pi}$ and $\phi(t)=\phi(0)+\frac{v^2t^2}{2\pi}$ with $v=0.01$. By extracting the non-Abelian quantum metric from full-time-dynamics simulations, we obtain the numerical results for the Chern number and Euler class as $\mathcal{C}{+}=-1.9988$ and $\chi{+}=0.9994$.

\section{Schemes for simulation of model Hamiltonians}
\label{sec6}
\begin{figure}
\centering
 \includegraphics[width=0.45\textwidth]{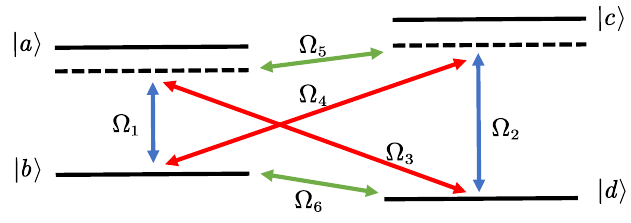}
\caption{Diagrammatic sketch of a four-level atomic system for simulating the Hamiltonian $H_{CP}^{(1)}$ and $H_{C_2T}^{(1)}$.}
\label{fig5}
\end{figure}
In this section, we first propose concrete experimental platforms for simulating the $CP$ symmetric and $C_{2}T$ symmetric Hamiltonians $H_{CP}^{(1)}$ and $H_{C_2T}^{(1)}$ with ultracold atoms in the parameter space, following the manipulation of four-level ${ }^{87} \mathrm{Rb}$ atoms in Ref. \cite{QXLv2021}. These Hamiltonians can be realized using a four-level atomic system as shown in Fig.~\ref{fig5}. In a ${ }^{87} \mathrm{Rb}$ atomic system, we can choose the following four atomic levels: $\left|a\right\rangle=\left|F=2, m_{F}=-1\right\rangle$, $\left|b\right\rangle=\left|F=1, m_{F}=-1\right\rangle$, $\left|c\right\rangle=\left|F=2, m_{F}=0\right\rangle$ and $\left|d\right\rangle=\left|F=1, m_{F}=0\right\rangle$. Using the bare state basis $\{ |a\rangle,|b\rangle,|c\rangle,|d\rangle    \}$, the Hamiltonian $H_{CP}^{(1)}$ is given by
\begin{equation}
\begin{aligned}
H^{\prime}_{CP}=&\omega_{a}\left|a\right\rangle\left\langle a\right|+\omega_{b}\left|b\right\rangle\left\langle b\right|+\omega_c\left|c\right\rangle\left\langle c\right|+\omega_d\left|d\right\rangle\left\langle d\right| \\
&+(\Omega_1 e^{i \omega_1 t} e^{i \varphi_1}|a\rangle\left\langle b\left|+\Omega_2 e^{i \omega_2 t} e^{i \varphi_2}\right| a\right\rangle\langle c| \\
&+\Omega_3 e^{i \omega_3 t} e^{i \varphi_3}|a\rangle\left\langle d\left|+\Omega_4 e^{i \omega_4 t} e^{i \varphi_4}\right| b\right\rangle\langle c| \\
&+\Omega_5 e^{i \omega_5 t} e^{i \varphi_5}|b\rangle\left\langle d\left|+\Omega_6 e^{i \omega_6 t} e^{i \varphi_6}\right| c\right\rangle\langle d| + H.c.),
\end{aligned}
\end{equation}
where $\omega_{i}$ $(i=a,b,c,d)$ are the energy frequencies of $\left|i\right\rangle$, and $\Omega_{l}$, $\omega_{l}$, $\varphi_{l}$ $(l=1,2,3,4)$ correspond to the Rabi frequencies, frequencies and phases of the controlling microwaves, respectively. We can tune the Hamiltonian to the reference frame to obtain effective Hamiltonian $\mathcal{H}^{\prime}_{CP}=U^{\dagger} H^{\prime} U+i\left(\partial_{t} U^{\dagger}\right) U,$ where $U=\left|a\right\rangle\left\langle a|+e^{-i \omega_{1} t}|b\right\rangle\left\langle b\right|+e^{-i \omega_{2} t}|c\rangle\langle c|+e^{-i \omega_{3} t}|d\rangle\langle d|$
\begin{equation}
\mathcal{H}^{\prime}_{CP}=\left(\begin{array}{cccc}
\Delta_1 & i\Omega_{1} & \Omega_{2} & \Omega_{3} \\
-i\Omega_{1} & \Delta_2 & \Omega_{4} & \Omega_{5} \\
\Omega_{2} & \Omega_{4} & \Delta_3 & i\Omega_{6} \\
\Omega_{3} & \Omega_{5} & -i\Omega_{6} & \Delta_4
\end{array}\right),
\end{equation}
where $\Delta_1=\omega_a$, $\Delta_2=\omega_{b}-\omega_{1}$, $\Delta_3=\omega_{c}-\omega_{2}$, $\Delta_4=\omega_{d}-\omega_{3}$, and  $\varphi_1=\varphi_6=\pi/2$, $\varphi_2=\varphi_3=\varphi_4=\varphi_5=0$. The Hamiltonian is time independent. The Hamiltonian in Eq.~(\ref{H_1}) can be derived  if we set $\Delta_1=\Delta_2=\Delta_3=\Delta_4=0$,  $\left\{\Omega_1,\Omega_2,\Omega_3,\Omega_4, \Omega_5, \Omega_6\right\}=\left\{d_y,d_z,d_x,d_x,-d_z,d_y\right\}$. On the other hand,  the corresponding parameterized Hamiltonian in spherical coordinates can be constructed if the parameters become $\Delta_1=\Delta_2= \Delta_3= \Delta_4=0$, $\left\{\Omega_1,\Omega_2,\Omega_3,\Omega_4, \Omega_5, \Omega_6\right\}=\left\{\sin\theta\cos\phi,\cos\theta,\sin\theta\sin\phi,\sin\theta\sin\phi,-\cos\theta,\sin\theta\cos\phi\right\}$.

Using the same scheme, here we present a concrete experimental platform to simulate the Hamiltonians with the form of Eq. (\ref{hct}). As shown in Fig.~\ref{fig5}, this Hamiltonian can be realized with the same four-level atomic system and $\Omega_5=\Omega_6=0$, which is given by
as
\begin{equation}
\begin{aligned}
H^{\prime}_{C_2T}=&\omega_{a}\left|a\right\rangle\left\langle a\right|+\omega_{b}\left|b\right\rangle\left\langle b\right|+\omega_c\left|c\right\rangle\left\langle c\right|+\omega_d\left|d\right\rangle\left\langle d\right| \\
&+\left(\Omega_{1}e^{i \omega_{1} t} e^{i \varphi_{1}}\left|a\right\rangle\left\langle b\right|\right.+\Omega_{2}e^{i \omega_{2} t} e^{i \varphi_{2}}\left|a\right\rangle\left\langle d\right| \\
&+\Omega_{3}e^{i \omega_{3} t} e^{i \varphi_{3}}\left|b\right\rangle\left\langle c\right| \left.+\Omega_{4}e^{i \omega_{4} t} e^{i \varphi_{4}}\left|c\right\rangle\left\langle d\right|+\text{H.c.}\right).
\end{aligned}
\end{equation}
Under the reference frame, we obtain the effective Hamiltonian       $\mathcal{H}^{\prime}_{C_2T}=U^{\dagger} H^{\prime} U+i\left(\partial_{t} U^{\dagger}\right) U,$ where $U=\left|a\right\rangle\left\langle a|+e^{-i \omega_{1} t}|b\right\rangle\left\langle b\right|+e^{-i (\omega_{1}+\omega_{3}) t}|c\rangle\langle c|+e^{-i(\omega_{1}+\omega_{3}+\omega_{4}) t} |d\rangle\langle d|$ as
\begin{equation}
\mathcal{H}^{\prime}_{C_2T}=\left(\begin{array}{cccc}
\Delta_{1} & \Omega_{1} & 0 & \Omega_{2} \\
\Omega_{1} & \Delta_{2} & \Omega_{3} & 0 \\
0 & \Omega_{3} & \Delta_{3} & \Omega_{4} \\
\Omega_{2} & 0 & \Omega_{4}  & \Delta_{4}
\end{array}\right),
\end{equation}
where $\Delta_1=\omega_a$, $\Delta_2=\omega_b-\omega_1$, $\Delta_3=\omega_c-\omega_1-\omega_3$, $\Delta_4=\omega_d-\omega_1-\omega_3-\omega_4$, and $\varphi_1=\varphi_2=\varphi_3=\varphi_4=0$. The Hamiltonian is time independent. The Hamiltonian in Eq.~(\ref{hct}) is achieved by setting $\left\{\Delta_1, \Delta_2, \Delta_3, \Delta_4\right\}=\left\{d_x,-d_x,d_x,-d_x\right\}$ and $\left\{\Omega_1,\Omega_2,\Omega_3,\Omega_4\right\}=\left\{d_z,-d_y,d_y,d_z\right\}$. The  parameterized Hamiltonian can be constructed for $\left\{\Delta_1, \Delta_2, \Delta_3, \Delta_4\right\}=\left\{\sin\theta\sin\phi,-\sin\theta\sin\phi,\sin\theta\sin\phi,-\sin\theta\sin\phi\right\}$ and $\left\{\Omega_1,\Omega_2,\Omega_3,\Omega_4\right\}=\left\{\cos\theta,-\sin \theta\cos \phi,\sin\theta\cos\phi,\cos\theta\right\}$.

 Lastly, we briefly discuss an experimental scheme of realizing the four-band Dirac Hamiltonians with ultracold atoms in optical lattices, based on the recent advances in synthetic gauge field and spin-orbit coupling for engineering topological phases \cite{DWZhang2018}. We take the Hamiltonian $H_{CP}^{(1)}$ in Eq.~(\ref{H_1}) as an example (the other Hamiltonian can also be realized in a similar way). To realize the Hamiltonian $H_{CP}^{(1)}$, one can use non-interacting fermionic atoms loaded in a three-dimensional optical lattice and choose four atomic internal states in the ground state manifold. Similar to the propose in Ref. \cite{JTWang2024}, one can choose four Zeeman-split ground hyperfine levels in ground state manifold $S_{1/2}$: $\left|e_{\uparrow,\downarrow}\right\rangle=\left|F+1, m_{F}=\pm 1\right\rangle$ and $\left|g_{\uparrow,\downarrow}\right\rangle=\left|F, m_{F}=\pm 1\right\rangle$. 
In real space, the corresponding tight-binding Hamiltonian is given by
\begin{equation}
\begin{aligned}
\hat{H}_{\mathbf{r}}&=\sum_{\mathbf{r}}(t\hat{H}_{\mathbf{x}}+t\hat{H}_{\mathbf{y}}+t\hat{H}_{\mathbf{z}}+\hat{H}_{M})+\text{H.c.}, \\
\hat{H}_{\mathbf{x}}&=\hat{a}_{e,\uparrow,\mathbf{r}}^{\dagger}(i\hat{a}_{g,\downarrow,\mathbf{r}-\mathbf{e}_x}\!-\!i\hat{a}_{g,\downarrow,\mathbf{r}+\mathbf{e}_x}\!-\!\hat{a}_{g,\uparrow,\mathbf{r}+\mathbf{e}_x}\!-\!\hat{a}_{g,\uparrow,\mathbf{r}-\mathbf{e}_x}) \\
&+\hat{a}_{e,\downarrow,\mathbf{r}}^{\dagger}(i\hat{a}_{g,\uparrow,\mathbf{r}-\mathbf{e}_x}\!-\!i\hat{a}_{g,\uparrow,\mathbf{r}+\mathbf{e}_x}\!+\!\hat{a}_{g,\downarrow,\mathbf{r}+\mathbf{e}_x}\!+\!\hat{a}_{g,\downarrow,\mathbf{r}-\mathbf{e}_x}), \\
\hat{H}_{\mathbf{y}}&=\hat{a}_{e,\uparrow,\mathbf{r}}^{\dagger}(\hat{a}_{e,\downarrow,\mathbf{r}-\mathbf{e}_y}-\hat{a}_{e,\downarrow,\mathbf{r}+\mathbf{e}_y}-\hat{a}_{g,\uparrow,\mathbf{r}+\mathbf{e}_y}-\hat{a}_{g,\uparrow,\mathbf{r}-\mathbf{e}_y}) \\
&+\hat{a}_{g,\downarrow,\mathbf{r}}^{\dagger}(\hat{a}_{g,\uparrow,\mathbf{r}+\mathbf{e}_y}-\hat{a}_{g,\uparrow,\mathbf{r}-\mathbf{e}_y}+\hat{a}_{e,\downarrow,\mathbf{r}+\mathbf{e}_y}+\hat{a}_{e,\downarrow,\mathbf{r}-\mathbf{e}_y}), \\
\hat{H}_{\mathbf{z}}&\!\!=\!\hat{a}_{g,\downarrow,\mathbf{r}}^{\dagger}(\hat{a}_{e,\downarrow,\mathbf{r}-\mathbf{e}_z}\!\!\!+\!\hat{a}_{e,\downarrow,\mathbf{r}+\mathbf{e}_z}\!)\!\!-\!\!\hat{a}_{e,\uparrow,\mathbf{r}}^{\dagger}(\hat{a}_{g,\uparrow,\mathbf{r}+\mathbf{e}_z}\!\!\!+\!\hat{a}_{g,\uparrow,\mathbf{r}-\mathbf{e}_z}\!), \\
\hat{H}_{M}&=M_{z}(\hat{a}_{e,\uparrow,\mathbf{r}}^{\dagger}\hat{a}_{g,\uparrow,\mathbf{r}}-\hat{a}_{e,\downarrow,\mathbf{r}}^{\dagger}\hat{a}_{g,\downarrow,\mathbf{r}}).
\end{aligned}
\end{equation}
Here $\hat{a}_{\tau,\sigma,\mathbf{r}}^{\dagger}$ $(\hat{a}_{\tau,\sigma,\mathbf{r}})$ is the fermionic creation (annihilation) operator at lattice site $\mathbf{r}$, with $\tau=\{e,g\}$ and $\sigma=\{\uparrow,\downarrow\}$, and $\mathbf{e}_{\mathbf{x}(\mathbf{y},\mathbf{z})}$ as the unit vector along the $\mathbf{x}(\mathbf{y},\mathbf{z})$ direction.
$\hat{H}_{\mathbf{rx(y,z)}}$ represents the hopping along $\mathbf{x(y,z)}$ direction, and $\hat{H}_M$ denotes an effective on-site Zeeman term. The hopping terms with synthetic gauge potentials and spin-orbit couplings can be realized by applying two-photon Raman coupling with the laser beams of proper configurations, similar to those configurations proposed in Refs. \cite{JTWang2024,STWang2014}. The term $\hat{H}_M$ can be achieved by applying a radio-frequency field or Raman beams for coupling proper atomic internal states \cite{JTWang2024}. Furthermore, the band structures of topological semi-metals can be detected by the Bragg spectroscopy or Bloch-Zener oscillations of ultracold atoms in optical lattices \cite{DWZhang2018}. The QGT and the related topological invariants can be measured and extracted from the tomography of Bloch bands, as experimentally demonstrated in Ref. \cite{CRYI2023}.

\section{Conclusion}
\label{sec7}
In summary, we have derived a general relation between the non-Abelian quantum metric and the unit Bloch vector in a generic Dirac Hamiltonian with degenerate bands. Additionally, we have established a relation between the non-Abelian quantum metric and the Berry (Euler) curvature, providing alternative methods for calculating topological Chern number (Euler class). We have presented and investigated two specific classes of Hamiltonians with the $CP$ and $C_2T$ symmetries. The topological invariants characterizing the phase transitions are the Chern number and Euler class, which are obtained through integration of the QGT over the parameter space. For the reduced 2D insulator phases, we have computed the corresponding Wilson loop. Through adiabatic perturbation theory, we have demonstrated the connection between the non-Abelian quantum metric and the energy fluctuation, as well as the connection between the Berry curvature and generalized force. These non-adiabatic effects can be used to extract all components of the QGT, as numerically demonstrated for the $CP$ and $C_2T$ symmetric Hamiltonians. We have also proposed experimentally feasible schemes for quantum simulation of the model Hamiltonians in ultracold atomic platforms.

\begin{acknowledgments}
	We thank  Prof. Gong Jiangbin for helpful discussions. The work is supported by the National Key Research and Development Program of China (Grant No. 2022YFA1405300), and the National Natural Science Foundation of China (Grants No. 12074180 and No. 12174126). Hai-Tao Ding also acknowledges financial support from the China
Scholarship Council.\end{acknowledgments}

\begin{appendix}
\section{Construction of global degenerate Hamiltonian}
\label{appa}
In this paper, we denote 16 Clifford matrices
\begin{equation}
\left\{\sigma_0, \sigma_x, \sigma_y, \sigma_z\right\} \otimes\left\{\tau_0, \tau_x, \tau_y, \tau_z\right\},
\end{equation}
and set
\begin{equation}
\begin{aligned}
\Gamma_0 &=\sigma_0 \tau_0, \Gamma_1=\sigma_z \tau_x, \Gamma_2=\sigma_0 \tau_y,  \Gamma_3=\sigma_0 \tau_z, \\
\Gamma_4 & =\sigma_x \tau_x, \Gamma_{5}=\sigma_y \tau_x, \Gamma_{12}=\sigma_z \tau_z, \Gamma_{13}=-\sigma_z \tau_y, \\
\Gamma_{14} &=\sigma_2 \tau_0, \Gamma_{15}=\sigma_x \tau_0, \Gamma_{23}=\sigma_0 \tau_x, \Gamma_{24}=-\sigma_x \tau_z, \\
\Gamma_{25} & =-\sigma_y \tau_z, \Gamma_{34}=-\sigma_x \tau_y, \Gamma_{35}=\sigma_y \tau_y, \Gamma_{45}=\sigma_z \tau_0,
\end{aligned}
\end{equation}
where
\begin{equation}
\Gamma_{ab}=\frac{1}{2i}[\Gamma_a,\Gamma_b].
\end{equation}
The matrices $\Gamma_i$ $(i=1,2,3,4,5)$ are the generators of the Clifford algebra. $\Gamma_i$ satisfy the anti-commutation relations
\begin{equation}
\left\{\Gamma_a, \Gamma_b \right\}=2\delta_{ab},
\end{equation}
where $a,b=\left\{1,2,3,4,5\right\}$. There are also some other anti-commutation relations
\begin{equation}
\begin{aligned}
\left\{\Gamma_{ab},\Gamma_{c}\right\}&=\epsilon_{abcde}\Gamma_{de},  \\
\left\{\Gamma_{ab},\Gamma_{cd}\right\}&=2\epsilon_{abcde}\Gamma_e +2\delta_{ac}\delta_{bd}-2\delta_{ad}\delta_{bc}.
\end{aligned}
\end{equation}
Here we show how to build a general 4-bands Hamiltonian with global two-fold degeneracy with 5 Dirac matrices $\Gamma_{i}$ and 10 commutators $\Gamma_{ij}$. The total Hamiltonian can be written as
\begin{equation}
\label{A6}
H(k)=\sum_{i=1}^{5} B_i(k) \Gamma_i+\sum_{i<j} B_{i j}(k) \Gamma_{i j}.
\end{equation}
$B_{i}(k)$ and $B_{ij}(k)$ are real functions. If such system is two-fold degenerate, $H(k)$ satisfies
\begin{equation}
\label{Hequ}
H(k)^2=f(k)I_4,
\end{equation}
where $f(k)$ is a function of parameter $k$, $I_4$ is $4\times 4$ identity matrix. The eigen-energies of the Hamiltonian $H(k)$ are $E_{\pm}=\pm \sqrt{f(k)}$.
From Eq.~\eqref{A6}, the $H(k)^2$ is
\begin{equation}
\begin{aligned}
H(k)^2&=(\sum_{i}B_{i}^2+\sum_{i<j}B_{ij}^2) I_4 +\sum_{\substack{i \\ k<l}}B_{i}B_{kl}\epsilon_{klimn}\Gamma_{mn} \\
&+\sum_{\substack{i<j \\ k<l,i\neq k, j\neq l}}B_{ij}B_{kl}\epsilon_{ijklm}\Gamma_{m}.
\end{aligned}
\end{equation}
To satisfy Eq.~\eqref{Hequ}, the last two terms at right hand side should equal to zero
\begin{equation}
\begin{aligned}
\sum_{\substack{i \\ k<l}}B_{i}B_{kl}\epsilon_{klimn}\Gamma_{mn}&=0,            \\
\sum_{\substack{i<j \\ k<l,i\neq k, j\neq l}}B_{ij}B_{kl}\epsilon_{ijklm}\Gamma_{m}&=0.
\end{aligned}
\end{equation}
It restricts the general Hamiltonian can only takes following two forms
\begin{equation}
\begin{aligned}
H(k)&=\sum_{i=1}^{5}B_{i}\Gamma_{i}, \\
H(k)&=B_{i}\Gamma_{i}+\sum_{j}B_{ij}\Gamma_{ij}.
\end{aligned}
\end{equation}

\section{Derivation of  Eq.~\eqref{gfre}}
\label{appc}
For two global degenerate eigenstates $\left|\psi_i\right\rangle$ $(i=1,2)$, the non-Abelian quantum geometric tensor is
\begin{equation}
Q_{\mu \nu}^{ii}=\sum_{m}\left\langle\partial_\mu \psi_i \mid \psi_m\right\rangle\left\langle \psi_m \mid \partial_\nu \psi_i\right\rangle.
\end{equation}
$\left|\psi_m\right\rangle$ are excited states. The trace of matrix $Q_{\mu\nu}$ is
\begin{equation}
\begin{aligned}
\text{Tr}(Q_{\mu\nu}) &= \sum_{i} Q_{\mu \nu}^{ii} \\
&=\frac{1}{4 E_{+}^2} \sum_{i} \sum_{m}\left\langle \psi_i\left|\partial_\mu H\right| \psi_m\right\rangle\left\langle \psi_m\left|\partial_\nu H\right| \psi_i\right\rangle.
\end{aligned}
\end{equation}
Here we use the relation: $\left\langle \psi_i \mid \partial_\mu \psi_m\right\rangle\left(E_m-E_i\right)=\left\langle \psi_i\left|\partial_\mu H\right| \psi_m\right\rangle$. The trace of $g_{\mu\nu}$ equals to
\begin{equation}
\text{Tr}(g_{\mu\nu})=(\text{Tr}(Q_{\mu\nu})+\text{Tr}(Q_{\nu\mu}))/2,
\end{equation}
where
\begin{equation}
\begin{aligned}
\text{Tr}(g_{\mu\nu}) &= \sum_{i} g_{\mu \nu}^{ii}  \\
& =\frac{\sum_{r=1}^3 \partial_\mu d_r(\mathbf{k}) \partial_\nu d_r(\mathbf{k})-\partial_\mu d \partial_\nu d}{2 d^2} \\
& =\frac{1}{2} \sum_{r=1}^3 \partial_\mu \hat{d}_r \partial_\nu \hat{d}_r.
\end{aligned}
\end{equation}
Here $\hat{d}_{r}=d_{r}(\boldsymbol\lambda)/d$. For Hamiltonian in Eq.~\eqref{H_gen}, we define matrix $G_{\mu\nu}$
\begin{equation}
\begin{aligned}
 G_{\mu\nu}& \equiv\left(\begin{array}{ccc}
2\text{Tr}(g_{\mu\mu}) & 2\text{Tr}(g_{\mu\nu})
\vspace{1ex}\\
2\text{Tr}(g_{\nu\mu}) & 2\text{Tr}(g_{\nu\nu})  \\
\end{array}\right)  \\
&= \left(\begin{array}{ccc}
\partial_\mu \hat{d}_r \partial_\mu \hat{d}_r & \partial_\mu \hat{d}_r \partial_\nu \hat{d}_r
\vspace{1ex}\\
\partial_\nu \hat{d}_r \partial_\mu \hat{d}_r & \partial_\nu \hat{d}_r \partial_\nu \hat{d}_r  \\
\end{array}\right),
\end{aligned}
\end{equation}
where Einstein summation convention is used for the index $r$. The determinant of $G_{\mu\nu}$ equals determinant of following matrix, say $G_{\mu\nu}^{\prime}$
\begin{equation}
 G_{\mu\nu}^{\prime}\equiv\left(\begin{array}{ccc}
1 & 0 &0
\vspace{1ex}\\
0 & 2\text{Tr}(g_{\mu\mu}) & 2\text{Tr}(g_{\mu\nu})
\vspace{1ex}\\
0 & 2\text{Tr}(g_{\nu\mu}) & 2\text{Tr}(g_{\nu\nu})
\end{array}\right).
\end{equation}
$\det(G_{\mu\nu})=\det(G_{\mu\nu}^{\prime})$, $G_{\mu\nu}^{\prime}$ can be decomposed into the product of a matrix, say $A$, and its transpose matrix $A^{T}$
\begin{equation}
\begin{aligned}
 G^{\prime}& \equiv AA^{T}  \\
&=\left(\begin{array}{ccc}
\hat{d}_{1} & \hat{d}_{2} & \hat{d}_{3}
\vspace{1ex}\\
\partial_\mu \hat{d}_1 & \partial_\mu \hat{d}_2 & \partial_\mu \hat{d}_3
\vspace{1ex}\\
\partial_\nu \hat{d}_1 & \partial_\nu \hat{d}_2 & \partial_\nu \hat{d}_3
\end{array}\right) \left(\begin{array}{ccc}
\hat{d}_{1} & \partial_\mu \hat{d}_1 & \partial_\nu \hat{d}_1
\vspace{1ex}\\
\hat{d}_{2} & \partial_\mu \hat{d}_2 & \partial_\nu \hat{d}_2
\vspace{1ex}\\
\hat{d}_{3} & \partial_\mu \hat{d}_3 & \partial_\nu \hat{d}_3
\end{array}\right).
\end{aligned}
\end{equation}
The determinant of the matrix $A$ is
\begin{equation}
\det A =\det A^{T}=\epsilon_{\alpha \beta \gamma} \hat{d}_\alpha \partial_{\mu} \hat{d}_\beta \partial_{\nu} \hat{d}_\gamma.
\end{equation}
Thus we have
\begin{equation}
\sqrt{\det G_{\mu\nu}}=|\epsilon_{\alpha \beta \gamma} \hat{d}_\alpha \partial_{\mu} \hat{d}_\beta \partial_{\nu} \hat{d}_\gamma|,
\end{equation}
which is just the Eq.~\eqref{gfre}.

\end{appendix}

\end{document}